\documentclass[
    aps,12pt,final,notitlepage,
    oneside,onecolumn,nofootinbib,noshowpacs,
    nolongbibliography,centertags
]{revtex4-2}

\usepackage[english]{babel}

\usepackage{amsmath}

\usepackage[]{graphicx}
\graphicspath{{figures/}}

\usepackage{natbib}

\usepackage[usenames,dvipsnames]{xcolor} % link colors tweak
\usepackage[
    colorlinks = true,
    urlcolor = MidnightBlue,
    linkcolor = Red,
    citecolor = NavyBlue
]{hyperref}

\newcommand{{\yeas}}{{YEASa}}
\newcommand{{\E}}{{E_0}}

\newcommand{{\ethr}}{\varepsilon_{\text{thr.}}} % threshold energy
\newcommand{{\lnA}}{\langle\ln{A}\rangle} % <lnA>
\newcommand{{\xmax}}{x_{\text{max}}}      % Xmax
\newcommand{{\xobs}}{x_{\text{obs}}}
\newcommand{{\dE}}{{\Delta E_S}}          % SD energy deposit
\newcommand{{\Xmp}}{x_{\text{max}}^p}
\newcommand{{\XmFe}}{x_{\text{max}}^{\text{Fe}}}
\newcommand{{\XmExp}}{x_{\text{max}}^{\text{exp}}}
\newcommand{{\rhom}}{\rho_{\mu}}          % muon density
\newcommand{{\rhos}}{\rho_s}              % surfase density
\newcommand{{\Ebin}}{{E_{\text{bin}}}}
\newcommand{{\Ethin}}{{E_{\text{thin}}}}
\newcommand{{\wmax}}{{w_{\text{max}}}}

\newcommand{{\qgs}}{{\textsc{qgsj}et01}}
\newcommand{{\qgsii}}{{\textsc{qgsj}et-II.04}}
\newcommand{{\epos}}{{\textsc{epos}}}
\newcommand{{\eposlhc}}{{\textsc{epos-lhc}}}
\newcommand{{\sibyll}}{{\textsc{sibyll}-2.1}}
\newcommand{{\fluka}}{{\textsc{fluka}}}
\newcommand{{\corsika}}{{\textsc{corsika}}}

\newcommand{{\usec}}{{~$\mu$s}}    % usec
\newcommand{{\sqrm}}{{~m$^2$}}     % m^2
\newcommand{{\sqrkm}}{{~km$^2$}}   % km^2
\newcommand{\degr}{^{\circ}}     % degrees

\begin{document}

\title{Content of muons in extensive air showers measured by Yakutsk array}

\author{A.\,V.~Glushkov}
\email{glushkov@ikfia.ysn.ru}

\author{K.\,G.~Lebedev}
\affiliation{Yu.~G.~Shafer Institute of Cosmophysical Research and Aeronomy SB RAS, \\
    31 Lenin Ave., 677027 Yakutsk, Russia}

\author{A.\,V.~Saburov}
\email{vs.tema@gmail.com}

\begin{abstract}
    Lateral distribution of muons in extensive air showers registered at the Yakutsk array was investigated. The analysis covers two periods of observation: before 2018 and after 2020, when the revision of muon detectors with 1-GeV threshold was complete. Measured values of muon density are compared to computational results obtained within the framework of two hadron interaction models. Within the energy domain above $10^{18}$~eV the best agreement is observed for proton composition primary cosmic rays.
\end{abstract}


\maketitle

\section{Introduction}

High energy muons in extensive air showers (EAS) reflect various properties of primary cosmic rays (CR) since their significant fraction that is registered at the seal level originates from hadrons produced during initial acts of air shower development. The HiRes-MIA experiment was first to report the reduced number of muons in computational results obtained with the use of high-energy hadron interaction models in relation to their data~\cite{AbuZayyad:PRL(2000)}. The effect was later confirmed by NEVOD-DECOR~\cite{Bogdanov:PAN(2010), Bogdanov:APP(2018), Yurina:BRASP(2019), Yurina:BRASP(2021)} and Auger~\cite{Aab:PRD:91(2015), Aab:PRL:117(2016)} experiments and was also revealed in the data of the SUGAR array~\cite{Bellido:PRD:98(2018)}. By that time it has been established as a partial consensus that neither of the available hadron interaction models, including their latest generation which was calibrated by the Large Hadron Collider (LHC) data, could correctly describe the muon component of EAS. Large meta-analysis performed by the international working group~\cite{WHISP(2019), WHISP(ICRC2019)} have shown that in several experiments theoretical predictions agree with measured values up to CR energies $10^{16}$~eV but start diverging at higher energies and the difference increases with energy~\cite{Abbasi:PRD:98(2018), Muller:EPJConf(1019)}.

On the other hand some experiments did not confirm this effect. No muon excess in relation to simulation results was found in archival data of EAS-MSU~\cite{Fomin:PRL:92(2017)} and KASCADE-Grande~\cite{Apel:APP(2017)} arrays. The Yakutsk EAS array also did not detect a divergence between measured values and theoretical predictions~\cite{Glushkov:JETPLett(2019)}. Of particular note is the IceCube observatory, which had observed the muon excess, but at significantly lower energies; and starting from $\sim 10^{17}$~eV computational results converge with their measurements~\cite{Gonzales:EPJConf(2019)}.

Recent analysis of the AGASA data has also revealed the increased number of muons in comparison with predictions of the latest generation of hadron interaction models~\cite{Gesualdi:PRD:101(2020)}. This problem obviously arises from combination of several factors, such as the sheer complexity of making a correct description for hadronic interactions occurring at ultra-high energies and peculiarities of measurement techniques and data interpretation adopted in different experiments.

\section{Mean lateral distribution of muons in EAS}

Here we analyze the density of muon fluxes in EASs with primary energy above $10^{17}$~eV registered at the Yakutsk EAS array (The Unique Scientific Facility ``The D.\,D.~Krasilnikov Yakutsk Complex EAS Array''~\cite{YEASA:USU}). The mean muon density was estimated at distances 300, 600 and 1000~m from shower axis with the use of the mean lateral distribution function (MLDF) technique.

At the Yakutsk array the muon component is registered with underground scintillation detectors with a thickness of 5~cm and area of 2~\sqrm. In air shower events they essentially measure energy deposit $\Delta\epsilon$ which is proportional to the number of particles passing through them. At distance $r$ from shower axis it is expressed in units of a scintillation detector response, $\epsilon_1$:

\begin{equation}
    \rho(r) = \frac{\Delta\epsilon(r)}{\epsilon_1}~\text{m}^{-1}\text{,}
    \label{eq:1}
\end{equation}

\noindent
where $\epsilon_1=11.75$~MeV is the energy deposited in a detector by a vertical relativistic muon (vertical equivalent muon~--- VEM). Detectors are constantly calibrated and adjusted by amplitude density spectra collected from background cosmic particles~\cite{Glushkov:SovSymp(1974)}.

The analysis covers the data collected by three underground points of muon registration located at distances $0.5-1.0$~km from the array center. Inside each point ten 2~\sqrm{} scintillation detectors are placed under a layer of ground and concrete, this layer forms a shield with a threshold about 1~GeV. Until 2019 the detectors in a point were combined in pairs and each pair was connected to a logarithmic $LC$-converter. After a major revision of equipment which was complete in 2021, each detector was connected to a separate logarithmic $RC$-converter.

To obtain the muon MLDF, events were selected with preliminary energy estimation above $10^{17}$~eV and with errors of axes location below 50~m. The whole set of selected events was divided into groups (bins) by values of estimated energy $E$ with logarithmic step $\Delta\lg(E/\text{eV}) = 0.2$. A refined energy estimation within each bin was calculated from MLDFs of array's surface-based detector response described in~\cite{Glushkov:BRASP(2019)}. This method utilizes the following calorimetric relation:

\begin{equation}
    E = (3.76 \pm 0.30) \times 10^{17} \cdot
    S(600, 0\degr)^{1.02 \pm 0.02}~\text{[eV] ,}
    \label{eq:2}
\end{equation}

\noindent
where $S(600, 0\degr)$ is the signal of surface-based scintillation detectors located at 600~m from the axis recalculated to vertical direction of a shower.

\begin{figure}
    \centering
    \includegraphics[width=0.85\textwidth]{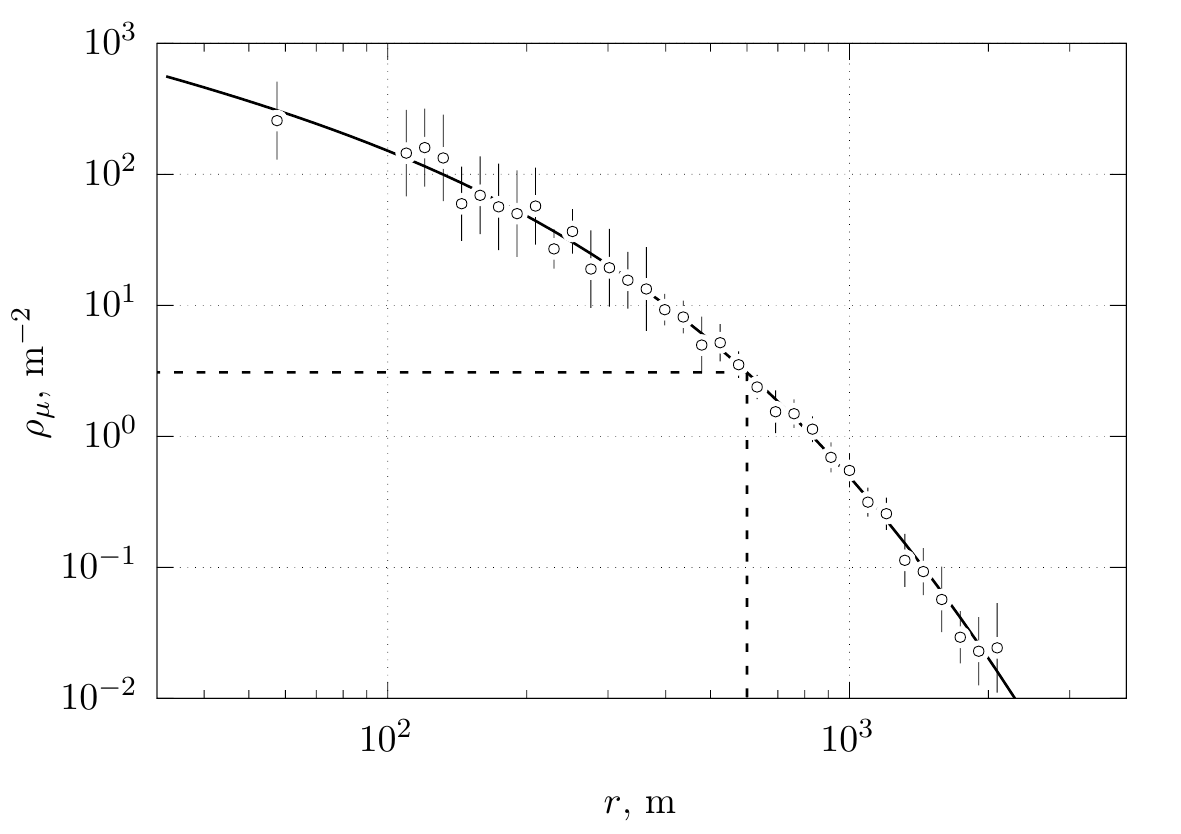}
    \caption{MLDF of muons in EASs with $\Ebin = 8.83 \times 10^{18}$~eV and $\left<\cos\theta\right> = 0.9$. Solid line~--- approximation (\ref{eq:4}) with parameters $\lg\left(\rhom(600,\theta)\right)=0.49 \pm 0.02$ (the value and axis distance $r=600$~m are marked with dashed lines) and $b=2.14 \pm 0.03$. The agreement criterion is $\chi^2=25.1$ for all data points.}
    \label{fig:1}
\end{figure}

To construct the MLDF of muon component, densities $\rhom(r_i)$ measured in individual events with energy $E$ were multiplied by a normalization ratio $\Ebin/E$, where $\Ebin$ was the mean energy of a group. Normalized densities were averaged out within intervals of axis distance resulting from logarithmic binning with $\Delta\lg{r} = 0.04$ step. Mean muon densities in intervals with logarithmic center at $\lg{r_i}$ were calculated as

\begin{equation}
    \left<\rhom\right> = 
    \frac{
        \sum{\rhom(r_i)}
    }{
        N_1 + N_0
    }\text{,}
    \label{eq:3}
\end{equation}

\noindent
where $N_1$ is the number of detectors fired within $i$-th radial interval, $N_0$~--- is the number of ``silent'' detectors within the interval, that haven't recorded a single particle while in data collecting mode. The resulting lateral profiles of muon density were approximated with the MLDF:

\begin{equation}
    \rhom(r,\theta) = \rhom(600, \theta) \cdot
    \left(\frac{600}{r}\right)^{0.75}
    \cdot
    \left(
        \frac{r_0 + 600}{r_0 + r}
    \right)^{b - 0.75}
    \cdot
    \left(
        \frac{2000 + 600}{2000 + r}
    \right)^{6.5}\text{,}
    \label{eq:4}
\end{equation}

\noindent
where $r_0=280$~m, $\rhom(600,\theta)$ and $b$ are free parameters calculated during the $\chi^2$-test~\cite{Glushkov:BRASP(2015)}. On fig.~\ref{fig:1} a mean lateral density distribution of muons is shown for a set of showers with zenith arrival directions $\theta<38\degr$, registered during the period before 2019~\cite{Glushkov:JETPLett(2019)}. Displayed errors

\begin{displaymath}
    \delta\rhom(r_i) =
    \frac{
        \sigma\rhom(r_i)
    }{
        \left<\rhom(r_i)\right>
    }
\end{displaymath}

\noindent
include the whole number of readings $N_1+N_0$ together with other multiple factors affecting the measurement and currently it is virtually impossible to distinguish and isolate them.

In works~\cite{Glushkov:PRD(2014), Sabourov:PhD(2018)} responses of surface-based and underground scintillation detectors of Yakutsk EAS array were calculated for events with energy above $10^{17}$~eV. This result was derived from the set of simulated showers generated with the use of \corsika-7.3700 code~\cite{CORSIKA:OG} within the framework of \qgs~\cite{QGSJet01} and \qgsii~\cite{QGSJetII04} hadron interaction models. For treatment of hadron interactions at energies below 80~GeV the \fluka{}2011 generator was used~\cite{FLUKA2011}. Showers were simulated within the energy range $10^{17}-10^{19.5}$~eV with logarithmic step $\Delta\lg(E/\text{eV})=0.5$ for primary protons ($p$) and iron nuclei (Fe), with zenith arrival angles $0\degr-60\degr$. In order to speed up the calculations the thin-sampling mechanism was activated with thinning level $\Ethin=(10^{-6}-10^{-5}$) and weight limit for all components $\wmax=E \cdot \Ethin$. For each set of primary parameters (primary particle, $E$, $\theta$) from 200 to 500 events were generated and based on this statistics a MLDF of muon detector response was obtained with radial logarithmic binning with $\Delta\lg(r)=0.04$ step.

\begin{figure}
    \centering
    \includegraphics[width=0.85\textwidth]{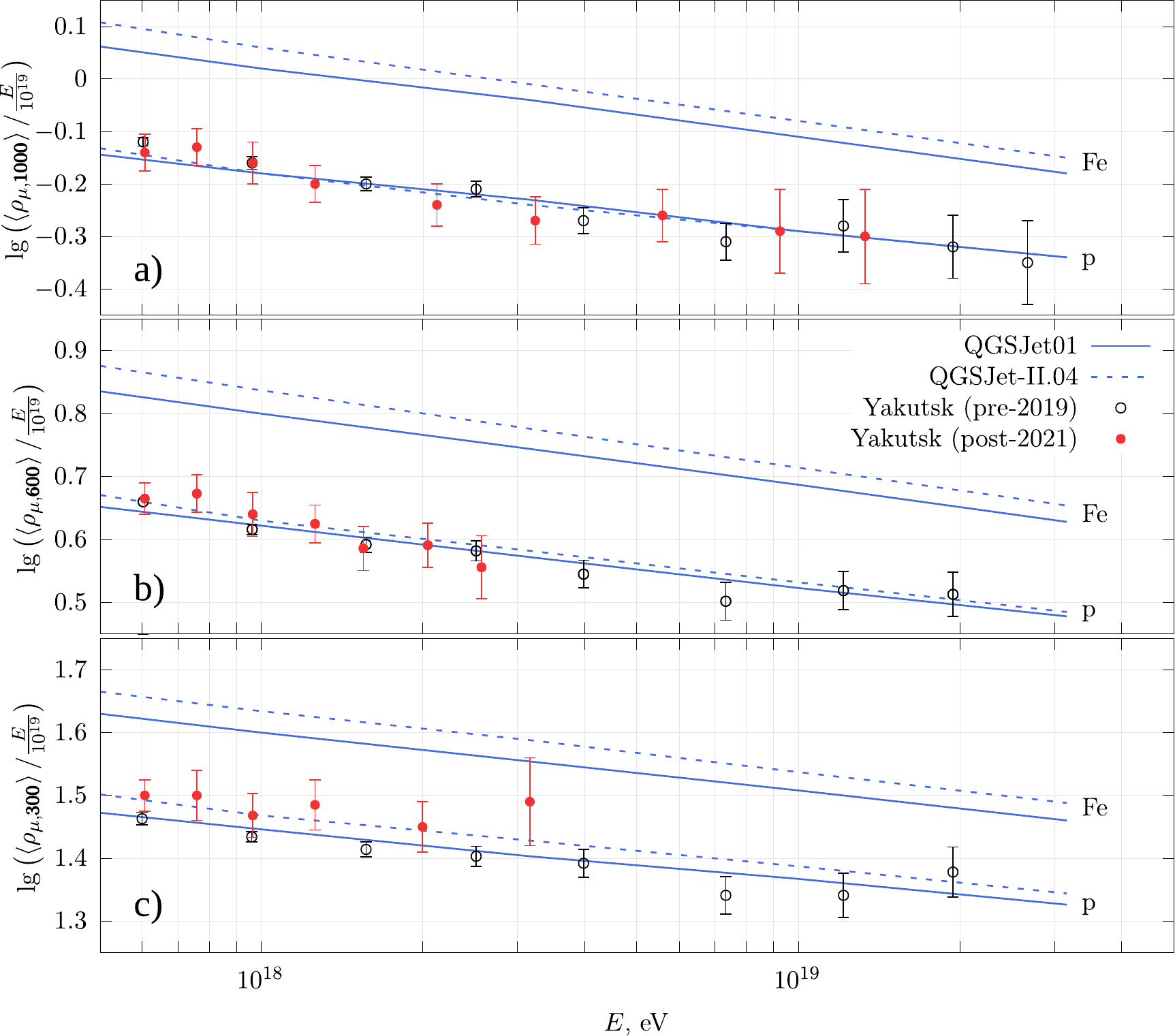}
    \caption{Muon densities at axis distances 1000~m (a), 600~m (b) in air showers with mean cosine of arrival zenith angles $\cos\theta \simeq 0.9$. Densities at 300~m from the axis (c) were measured earlier~\cite{Glushkov:JETPLett(2019)}. Lines represent predicted values obtained within the frameworks of \qgs{} (solid) and \qgsii{} (dashed) models.}
    \label{fig:2}
\end{figure}

On fig.\ref{fig:2} muon densities are shown for distances 1000, 600 and 300~m from shower axis, obtained from the approximation (\ref{eq:4}) and multiplied by normalization factor $E/10^{19}$. Different symbols denote two periods of the array operation, before 2019 and after 2021. Within statistical errors both data sets are in a good agreement with each other. The results of calculations performed within the frameworks of \qgs{} and \qgsii{} models are shown with lines. For both chosen models on all three axis distances measured muon densities agree with proton cosmic ray composition within energy range $(10^{17.7}-10^{19.3})$~eV.

We consider this result as a preliminary. Its further testing and verification will depend on the increase of operation statistics of reconfigured muon detectors and on refinement of their amplitude calibration.

\section{Conclusions}

Measurement of muon densities in EASs at 600 and 1000~m from the axis performed at the Yakutsk array have confirmed our earlier conclusion that cosmic rays with energies above $10^{18}$~eV are mostly protons. It was based on the analisys of muon density at 300~m from the axis. On all three distances there is no observable muon excess in EASs in relation to model predictions.

\section*{Funding}
This work was made within the framework of the state assignment No. 122011800084-7.

\begin{acknowledgments}
    Authors would like to express their gratitude to the staff of the Separate structural unit of ShICRA SB RAS ``The D. D. Krasilnikov Yakutsk complex EAS Array''.
\end{acknowledgments}

\section*{Conflict of interest}

The authors declare that they have no conflicts of interest.

\bibliography{muon-content-ykt2022}

\end{document}